# Soonspot: Software to Determine Areas and Sunspot Positions

P. Galaviz[1] • V.M.S. Carrasco[2,3] • F. Sánchez-Bajo[4] • M.C. Gallego[2,3] • J.M. Vaquero[1,3]

[1] Departamento de Física, Universidad de Extremadura, 06800 Mérida, Spain

[2] Departamento de Física, Universidad de Extremadura, 06006 Badajoz, Spain [e-mail: vmscarrasco@unex.es]

[3] Instituto Universitario de Investigación del Agua, Cambio Climático y Sostenibilidad (IACYS), Universidad de Extremadura, 06006 Badajoz, Spain

[4] Departamento de Física Aplicada, Universidad de Extremadura, 06006 Badajoz, Spain

**Abstract:** A new software (Soonspot) for the determination of the heliographic coordinates and areas of sunspots from solar images is presented. This program is very user-friendly and the accuracy of its results has been checked by using solar images provided by the Debrecen Photoheliographic Data (DPD). Due to its applicability in the studies of historical solar observations, the program has been used to analyze the solar drawings carried out by Hevelius in the 17th century.

## 1. Introduction

The Sun has been observed and studied by humans for millennia due to its influence in our lives (Vaquero and Vázquez, 2009). In particular, we have sunspot records available before the telescopic era, carried out mainly in Asia (Yau and Stephenson, 1988; Vaquero, Gallego, and García, 2002). However, the first sunspot drawing was made by a monk from Worcester (England) on 8th December in the year AD 1128 during the period known as the Medieval Solar Maximum and includes two sunspots (Willis and Stephenson, 2001). However, more or less systematic sunspot observations started with the use of the telescope as an astronomical instrument at the beginning of the 17th century (Hoyt and Schatten, 1998; Vaquero *et al*., 2016; Muñoz-Jaramillo and Vaquero, 2019).

The most striking sunspot drawings made in the first years of the telescopic era were carried out by Galileo Galilei, Christoph Scheiner, and Johannes Hevelius (Galilei and Scheiner, 2010; Hevelius, 1647). Since then, hundreds of observers have made drawings and taken photographs or digital images of the photosphere in order to study the phenomena that take place in it. Sunspot positions can be measured from these drawings or images and provide important information about the solar activity. For example, Ribes and Nesme-Ribes (1993) showed a strong hemispheric asymmetry during the Maunder Minimum from the calculation of the heliographic latitudes of the sunspots recorded by several observers of that time. More recently, Arlt (2009) presented the first butterfly diagram for 18$^{th}$ century from the Staudacher's sunspot records. We can also highlight the sunspot positions measured from Galileo's (Vokhmyanin and Zolotova, 2018) and Horrebow's (Karoff *et al*., 2019) sunspot records. Moreover, sunspots have been extensively used to calculate the solar rotation rates during past centuries basing on the determination of their heliographic coordinates (Abarbanell and Wöhl, 1981; Sánchez-Bajo *et al*., 2010). In addition, the sunspot area is a main indicator of the solar activity, closely related with other indices as the sunspot number (Carrasco *et al*., 2016). For this reason, accurate measurements of sunspot parameters (positions, areas, *etc*.) made from historical observations have a great interest.

The calculation of these sunspot parameters from drawings or photographs in historical archives is not a simple task, mainly due to the different instruments, procedures, *etc*., used by historical solar observers. In this way, the modern computer resources are especially powerful to analyze historical drawings. Regarding this fact, different computer programs have been proposed to measure sunspot parameters from digitized images. Among these programs, Sungrabber (Hrzina *et al*., 2007) has been used to analyze historical sunspot drawings made in 1884 at the Royal Astronomical Observatory of the Spanish Navy (Galaviz *et al*., 2016). In this way, drawings recorded by Zucconi in the period 1754-1760 were analyzed by using HSUNSPOTS (Cristo *et al*., 2011). More recently, CAMS (Computer Aided Measurement for Sunspots) has been proposed (Çakmak, 2014) to determine the heliographic coordinates of sunspot groups besides other parameters as, for example, the group length. In addition, Barata *et al*. (2018) developed a software to detect solar plages and Carrasco *et al*. (2018) another one to identify the sunspot umbrae and penumbrae in order to study the umbrae-penumbrae ratio during the Maunder Minimum. Although these programs provide good results in the measurement of the sunspot parameters, note that this field is open to enhance not only the technical aspects in the evaluation of them, but also other related with the appearance, flexibility and user-friendliness. In relation to this, in this work, we present the software Soonspot, developed with the aim of improving the functions and manageability presented in previous software of this kind. The outline of the paper is as follows: Section 2 is devoted to the presentation of the main features of the program, including a short guide of use. In Section 3, in order to check the accuracy of the results obtained with Soonspot, we compare the measured sunspot positions and areas of groups included in several solar images taken by the DPD (http://fenyi.solarobs.csfk.mta.hu/) with the data provided by this observatory about these same sunspot groups. Due to its main applicability to the analysis of historical solar observations, we have used Soonspot to analyze sunspot observations made by Hevelius (1647) in Section 4. Finally, the main conclusions of this work are presented in Section 5.

## 2. Description of the Software

Soonspot is a software intended to measure sunspot positions and areas from solar images. It has been programmed in M, the programming language of Matlab (https://www.mathworks.com) and it runs under the Windows operating system. We have used the user interface of Matlab to develop the software. In this way, the user does not have to know the programming code to work with the software and the application is easy to run. The requirements of the computer where the software will be installed are the same as for Matlab since it needs its routines (Matlab Component Runtime) to execute the application. These routines are incorporated in the install setup of the software. We note that the user does not need to install Matlab in their computer, only the Matlab Runtime in order to run the program. Furthermore, the install setup contains the executable file for Soonspot and a "readme" file including information about the installation and use of the program. The software can be downloaded from the website of the Historical Archive of Sunspot Observations (http://haso.unex.es/).

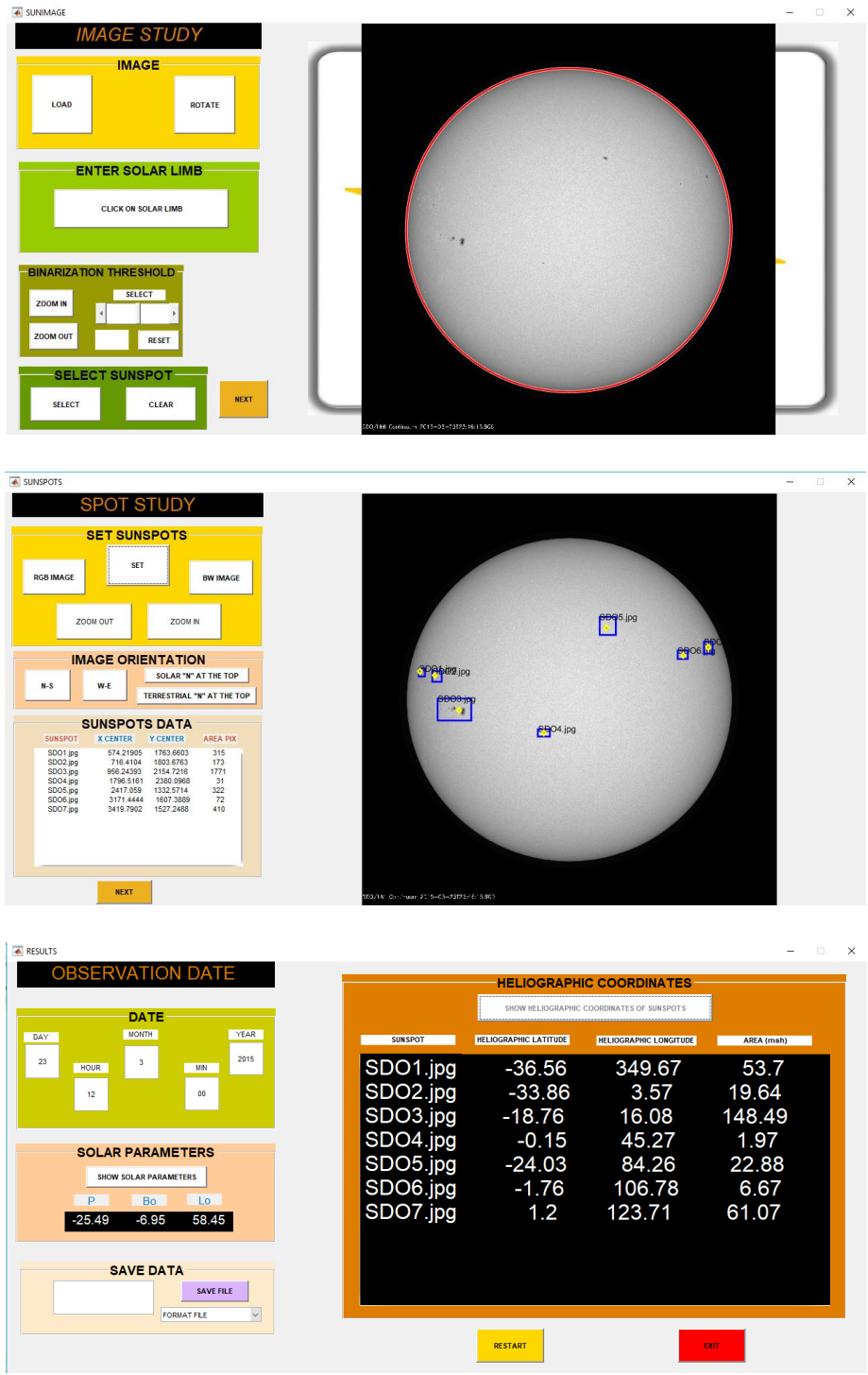

Figure 1. First (top panel), second (middle panel) and third (bottom panel) interface of Soonspot including a sunspot observation registered by telescope on board of the *Solar Dynamic Observatory* on 23 March 2015 [Source: https://sdo.gsfc.nasa.gov/].

In order to determine the sunspot parameters, users must complete three successive steps with different interfaces. In the first interface (Figure 1, top panel), the solar image must be loaded, the solar limb set and the image binarized. Soonspot can work with several image formats: jpg, tif or tiff and bmp. Moreover, once loaded, the image can be rotated in minimum intervals equal to 0.01°, both clockwise and counterclockwise, according to the needs of the user, without quality losses. A particular advantage of Soonspot over other programs is the ease to fix the solar limb, that is set by clicking in three different points of it, doing finally double click to finish the process. Thus, this task is done independently of the size, illumination or contrast of the image. We note that the solar disks in some historical drawings have an elliptical shape rather than circular and, in that case, the solar limb should be converted to a circle before using Soonspot in order to improve the accurate of the measurements. Once the solar limb is set, the following step involves the binarization of the image, necessary prior to calculate the sunspot areas. In regard to this, the user must select the binarization threshold. Note that the binarization process convert the RGB image (each pixel has a value between 0 and 255 as a function of its color) to a black and white image (each pixel can only have a value equal to 1 if the pixel is white or 0 if black). In this way, the binarization threshold ranges between 0 and 1 and all the pixels below that threshold are considered like black whereas those above the threshold are taken as white. According to this procedure, one can see that higher values of the binarization threshold imply darker binarized images. Once completed the binarization process, it is easy to calculate the area of a sunspot, because this is determined by the number of black pixels corresponding to that sunspot. In addition, as the umbra is darker than penumbra, the user can also measure only the umbral area applying a higher value of the binarization threshold. On the first interface, the user can zoom in any region of interest in order to choose the more suitable threshold. Figure 2 shows an example of binarization. Subsequently, the user must select the sunspots to study clicking the select button, delimiting the area of the sunspot with the mouse pointer and doing double click to save the data (each sunspot is independently saved). When all the sunspots are selected, the user must click on "Next" to pass to the second interface.

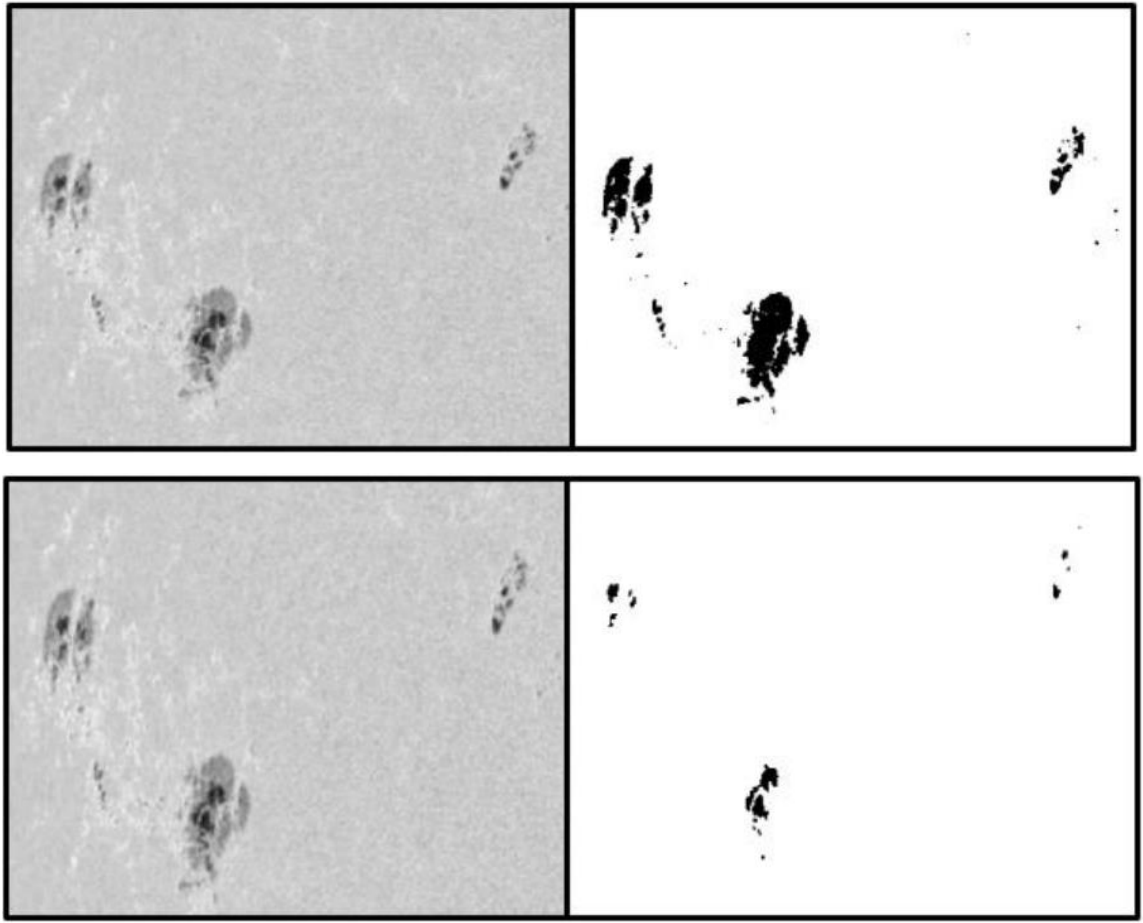

Figure 2. Results of the binarization process applied to sunspot 12420 observed on 22 September 2015 (from the Debrecen Photoheliographic Data). The binarization thresholds used were 0.7 (top panel) and 0.4 (bottom panel). Note that the area of the entire sunspot is reproduced in the top panel and only the umbral area of the sunspot in the bottom panel.

The selection of the sunspots can be confirmed in the second interface (Figure 1, middle panel) pressing the button "Set". In this second step, it can be also selected the orientation of the image with respect to the directions north-south and west-east by a double click on the north and west so that the first of two clicks is further north and west than the second one, respectively. The software also allows to select the north-south orientation of the image following the Earth or Sun axis. Note that the solar north and west of the image included in Figure 1 are located above and to the right of the image, respectively. Data about the sunspots like the name, area and position (or gravity

Soonspot: Software to Determine Areas and Sunspot Positions

center if an entire group was selected) in pixel units are displayed at the bottom left in the middle panel.

In order to determine heliographic coordinates, we need the observation date and the solar parameters *P*, $B_0$ and $L_0$. The observation date (in Gregorian calendar) will be introduced in the third interface, with the hour and minutes expressed in Universal Time. Note that these data can be changed avoiding to select newly the sunspots. Data about sunspots including names, areas (in millionths of solar hemisphere, msh) and heliographic coordinates are shown clicking on "Show Heliographic Coordinates of sunspot". These data can be saved in a .txt or .xlsx file pressing the button "Save".

The calculations of the coordinates are carried out using the formulae included in Meeus (1991). Note that the heliographic coordinates are corrected by precession and nutation and the area corrected by foreshortening is (Meadows, 2002):

$$A_\mathrm{M} = \frac{10^6 A_\mathrm{S}}{2\pi R^2 \cos \rho} = \frac{10^6 A_\mathrm{S}}{2\pi R^2 [\sin B_0 \sin B + \cos B_0 \cos B \cos(L - L_0)]}$$

where $A_M$ is the sunspot area in millionths of solar hemisphere; $A_S$, the sunspot area (in pixels); *R*, the radius of the solar image (in pixels); $\rho$, the angular distance from the solar disc center to the sunspot; *B*, the heliographic latitude of the sunspot; $B_0$, the heliographic latitude of the solar disc center; *L*, the heliographic longitude of the sunspot; and $L_0$, the heliographic longitude of the solar disc center.

3. A Comparison Using Data Recorded at the Debrecen Photoheliographic Data

In order to check the accuracy of the results obtained by Soonspot, we have compared the areas and heliographic coordinates calculated from Soonspot for sunspot groups recorded in several solar images obtained at the DPD (http://fenyi.solarobs.csfk.mta.hu/) with the positions and areas provided by DPD corresponding to the same groups (Table 1). We have selected one observation day per year from 2011 to 2017: 26 November 2011, 15 October 2012, 3 January 2013, 1 March 2014, 22 September 2015, 10 May 2016, and 4 September 2017. The total area of a group is calculated adding the area of all the sunspots of the group and the heliographic coordinates like the average between the sunspots that depicts the limits of each group. The mean error of the difference in the heliographic latitude and longitude measurements corresponding to the same groups obtained by both procedures is 0.14º in latitude and 0.23º in longitude. Furthermore, the maximum difference found in latitude is equal to 0.35º, corresponding to the group named by the DPD as 12415 on 22 September 2015 and 0.69º in longitude, corresponding to the group 11356 recorded by the DPD on 26 November 2011. We have obtained both mean relative errors in latitude and longitude lower than 1 %. These results show that Soonspot is an accurate tool to measure heliographic coordinates. Note, for example, that in the analysis of historical drawings, the uncertainty in the positions (unavailable) is the main source of errors and, for this reason, the uncertainties actually measure the precision of the data rather than its accuracy. Thus, we can assume that the errors in the determination of heliographic coordinates are around, in the worst case, one degree. In regard to the comparison of areas, the mean relative errors of the differences in the umbra and total area measurements between both procedures are greater than those obtained in the heliographic coordinates, that is, 23.3 % corresponding to total area and 44.6 % in the case of the umbral area. In general, the umbral area determined using Soonspot has been underestimated. However, in the case of the total areas, we have overestimated the measurements of the sunspot groups recorded on 26 November 2011, 15 October 2012 and 1 March 2014, whereas total areas were underestimated in the other four cases. These differences are probably due to the different procedures used to determine the sunspot areas, taking into account that the selected binarization threshold is the key parameter to provide the value of the umbral and total areas. Changes in this threshold lead to noticeable changes in these parameters.

Table 1. Comparison between areas and heliographic coordinates obtained from Soonspot and DPD for seven observation days. "Group" column shows the name assigned to each group, "Umbrae" and "Total" give the umbrae and total area (msh) measured according to both procedures, and "B" and "L" the latitude and Carrington longitude, respectively.

|            | Soonspot | | | | | DPD | | | | |
|------------|-------|--------|-------|-------|--------|--------|-------|-------|--------|-------|
|            | Group | Umbrae | Total | B     | L      | Umbrae | Total | B     | L      | Group |
| 26/11/2011 | A     | 35     | 471   | 19.49 | 192.01 | 52     | 324   | 19.58 | 192.02 | 11358 |
|            | B     | 3      | 53    | 17.84 | 213.70 | 11     | 36    | 17.98 | 213.5  | 11360 |
|            | C     | 21     | 217   | 17.54 | 223.08 | 24     | 158   | 17.72 | 223.04 | 11360 |
|            | D     | 33     | 185   | 15.25 | 233.75 | 27     | 180   | 15.57 | 233.06 | 11356 |

|  | E | 25 | 165 | 14.20 | 251.99 | 22 | 148 | 14.43 | 251.84 | 11355 |
|---|---|---|---|---|---|---|---|---|---|---|
|  | F | 0 | 81 | -23.87 | 283.30 | 15 | 66 | -23.86 | 283.74 | 11352 |
| 15/10/2012 | G | 63 | 361 | 7.45 | 228.79 | 69 | 350 | 7.35 | 228.64 | 11591 |
|  | H | 0 | 20 | 23.01 | 255.59 | 5 | 16 | 23.17 | 254.93 | 11592 |
|  | I | 27 | 300 | 12.85 | 266.86 | 36 | 272 | 12.55 | 266.78 | 11589 |
|  | J | 6 | 48 | -29.66 | 255.93 | 12 | 45 | -29.94 | 255.96 | 11590 |
|  | K | 11 | 106 | -12.55 | 302.81 | 17 | 93 | -12.68 | 302.69 | 11596 |
| 03/01/2013 | L | 0 | 18 | 13.89 | 230.13 | 11 | 78 | 13.88 | 230.03 | 11646 |
|  | M | 0 | 34 | 15.26 | 239.16 | 20 | 63 | 15.28 | 239.03 | 11644 |
|  | N | 37 | 130 | -23.46 | 252.04 | 35 | 164 | -23.56 | 251.99 | 11642 |
|  | O | 9 | 47 | 3.52 | 279.93 | 10 | 57 | 3.43 | 279.51 | 11641 |
|  | P | 4 | 51 | -13.05 | 290.96 | 11 | 63 | -13.02 | 290.8 | 11645 |
|  | Q | 28 | 115 | 12.75 | 308.41 | 23 | 155 | 12.88 | 308.3 | 11638 |
|  | R | 39 | 348 | 27.65 | 323.41 | 58 | 495 | 27.53 | 323.74 | 11640 |
| 01/03/2014 | S | 83 | 503 | -24.35 | 92.9 | 75 | 455 | -24.32 | 92.33 | 11991 |
|  | T | 79 | 447 | -14.28 | 108.94 | 61 | 384 | -13.94 | 108.87 | 11990 |
|  | U | 13 | 167 | 15.76 | 101.48 | 31 | 147 | 15.84 | 101.54 | 11993 |
|  | V | 6 | 24 | -19.42 | 139.64 | 6 | 22 | -19.62 | 139.67 | 11992 |
|  | W | 7 | 52 | -7.91 | 151.71 | 8 | 47 | -7.74 | 151.05 | 11994 |
|  | X | 30 | 232 | -1.35 | 152.13 | 33 | 192 | -1.27 | 152.05 | 11987 |
|  | Y | 1 | 154 | -10.27 | 186.58 | 34 | 173 | -10.31 | 187.17 | 11988 |
| 22/09/2015 | Z | 122 | 439 | 10.61 | 103.2 | 142 | 552 | 10.44 | 102.9 | 12420 |
|  | Aa | 7 | 47 | 15.42 | 120.07 | 23 | 56 | 15.16 | 120.11 | 12421 |
|  | Ab | 52 | 279 | -15.37 | 201.25 | 57 | 318 | -15.68 | 200.67 | 12418 |
|  | Ac | 21 | 194 | -19.08 | 235.78 | 48 | 276 | -19.43 | 235.6 | 12415 |
| 10/05/2016 | Ad | 4 | 49 | 21.18 | 300.3 | 28 | 84 | 21.01 | 300.22 | 12544 |
|  | Ae | 5 | 17 | -20.73 | 342.32 | 8 | 20 | -20.78 | 342.26 | 15445 |
|  | Af | 41 | 202 | 11.48 | 356.22 | 59 | 254 | 11.49 | 356.37 | 12542 |
|  | Ag | 19 | 80 | -5.3 | 0.75 | 36 | 87 | -5.34 | 0.66 | 12543 |
| 04/09/2017 | Ah | 0 | 19 | 17.72 | 49.57 | 17 | 51 | 17.76 | 49.56 | 12677 |
|  | Ai | 196 | 1043 | 13.2 | 104.09 | 245 | 1183 | 13.45 | 103.55 | 12674 |
|  | Aj | 182 | 856 | -8.96 | 117.85 | 253 | 922 | -8.92 | 117.82 | 12673 |
|  | Ak | 8 | 55 | -10.17 | 165.88 | 52 | 83 | -10.1 | 165.85 | 12676 |
|  | Al | 3 | 46 | -5.02 | 176.84 | 27 | 96 | -5.14 | 177.45 | 12675 |

**4. An Example: the Sunspot Observations Recorded by Hevelius in the 17th Century**

We have applied Soonspot to the sunspot observations recorded by Hevelius (1647) as an application example to historical observers. Hevelius made very detailed sunspot observations from the end of the year 1642 to the beginning of 1645. These observations have a great interest because they are the only systematic sunspot observations available just on the onset of the Maunder Minimum (Carrasco *et al*., 2019). Hevelius (1647) published these sunspot records in an appendix of the documentary source *Selenographia* providing both textual descriptions and sunspot drawings. Thus, we calculated the sunspot positions and areas recorded by Hevelius (1647).

Hevelius set a fixed line for the ecliptic which represents the east-west line in the drawings and therefore we have to correct the orientation of the images. First, the orientation of the images must be corrected according to the angle of the ecliptic with respect to the east-west direction of the celestial equator. Second, a sunspot observed in different days is represented in the same drawing. Note that the ecliptic fixed line is not referred to a particular observation day and that fact implies that the position measurements can be affected by a small error due to the daily variation of the orientation of the ecliptic (0.26º in average). By this reason, after correcting the ecliptic angle with respect to the east-west line, a procedure based on the minimization of the latitude variations of the sunspot groups has been used to calculate the heliographic coordinates. These corrections are explained in more detail in Galaviz (2018).

Sunspot positions and areas were measured after carrying out those corrections. The binarization threshold used for the calculations of the total and umbral areas was 0.7 and 0.35, respectively. Figure 3 depicts the sunspot positions calculated for all sunspots recorded by Hevelius (1647). Colors are according to the different area ranges established where dark colors represent the greatest sunspots. We can see that sunspots were observed in the interval ±20º in latitude and slightly appeared more in the northern solar hemisphere (541) than the southern (435) (Carrasco *et al*., 2019). Moreover, we have calculated the average of the umbra-penumbra ratio for all the sunspots obtaining a value equal to 0.26. This value agrees with works carried out in order to study this ratio both during the Maunder Minimum (Carrasco *et al*., 2018) and other epochs (Steinneger, 1990; Vaquero *et al*., 2005).

Soonspot: Software to Determine Areas and Sunspot Positions

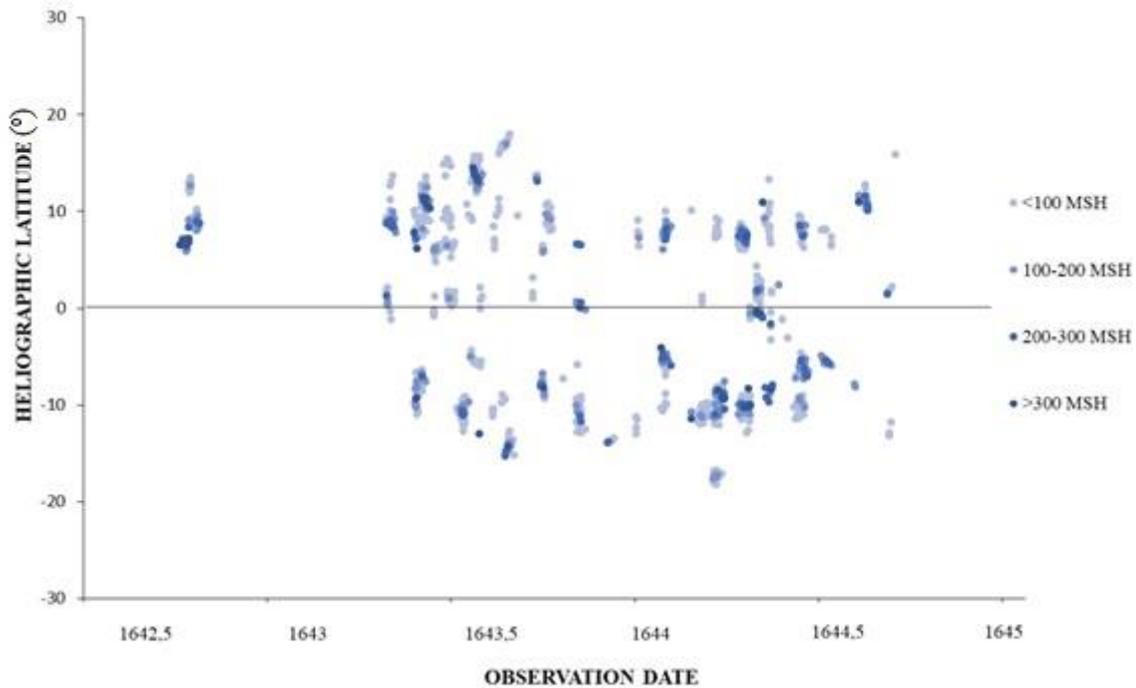

Figure 3. Sunspot positions according to the sunspot observations recorded by Hevelius (1647). Colors represent different area ranges with dark colors for the greatest areas.

In order to study the distribution of areas, we have obtained the relative size distribution of the umbral and total group areas recorded by Hevelius (1647) (Figure 4). We have followed the methodology employed by Bogdan *et al*. (1988). Thus, we can see that both the umbral and total group areas are found to be distributed lognormally such as is shown by Bogdan *et al*. (1988). We can see that lower areas present a greater deviation with respect to the lognormal distribution than larger areas. This fact can be due to a greater uncertainty in the determination of the lower sunspot areas. On the other hand, we show in Figure 5 how the selection in Soonspot of different binarization thresholds influences in the determination of areas (top panel), heliographic latitudes (middle panel), and longitudes (bottom panel). For that purpose, we have selected the sunspot observation made at DPD on 15 October 2012 where groups G, H, I, J, and K were recorded. According to the group areas, the binarization threshold that should be applied in Soonspot in order to obtain the area values provided by DPD is around 0.7 for all the groups. Note that the vertical grey bar in Figure 5 (top panel) includes all the cut-off between the curves that represent the area calculated from Soonspot applying different binarization thresholds and the area value provided by DPD (represented by the dashed horizontal lines). Furthermore, we can see that the binarization threshold does not provide significant errors in the determination of the heliographic latitudes and longitudes. The greatest differences in the calculation of the latitude and longitude are found in group G for latitudes and H for longitudes. Averaging the values of the latitudes and longitudes obtained from Soonspot for groups G and H applying different thresholds, we obtained a difference equal to 2.5% and 0.8 % with respect to the value provided by DPD, respectively.

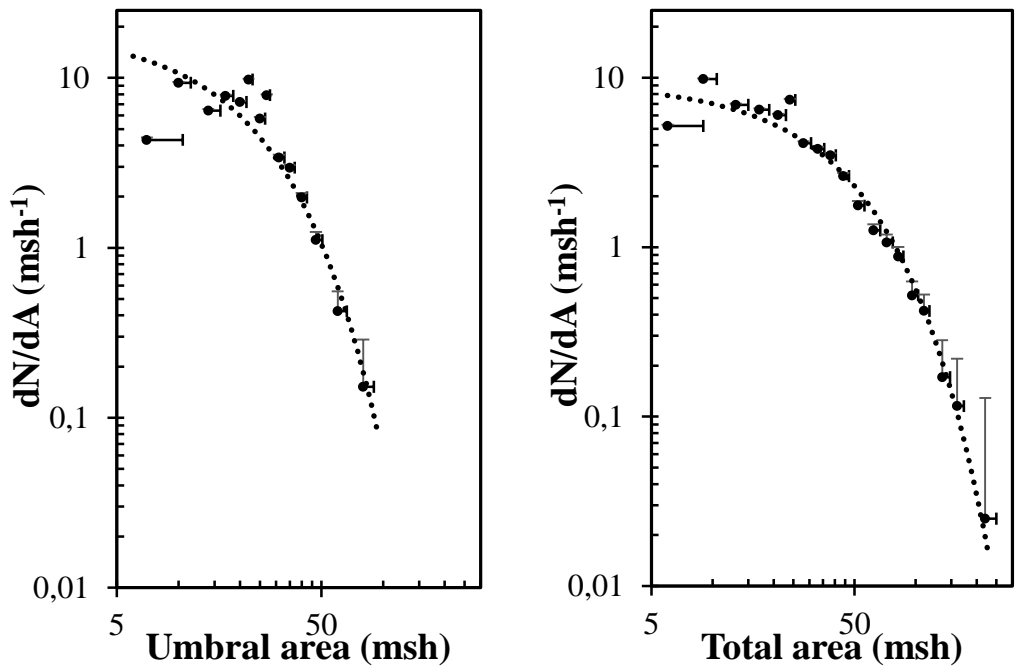

Figure 4. Umbral (left panel) and total (right panel) area distribution according to the sunspot observation recorded by Hevelius (1647). The areas are given in millionths of solar hemisphere (msh).

Soonspot: Software to Determine Areas and Sunspot Positions

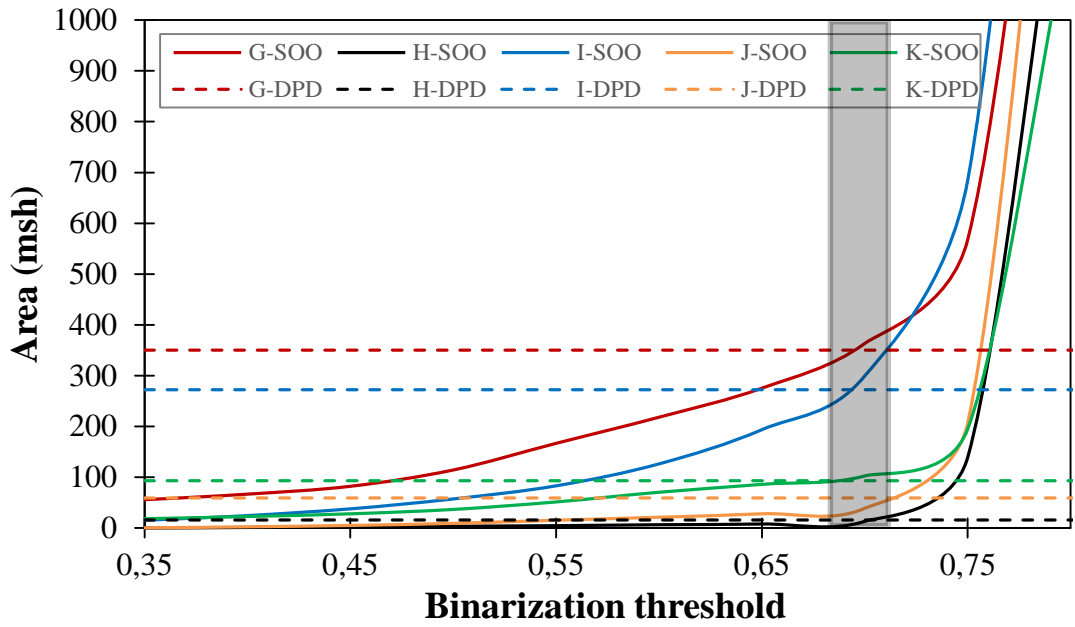

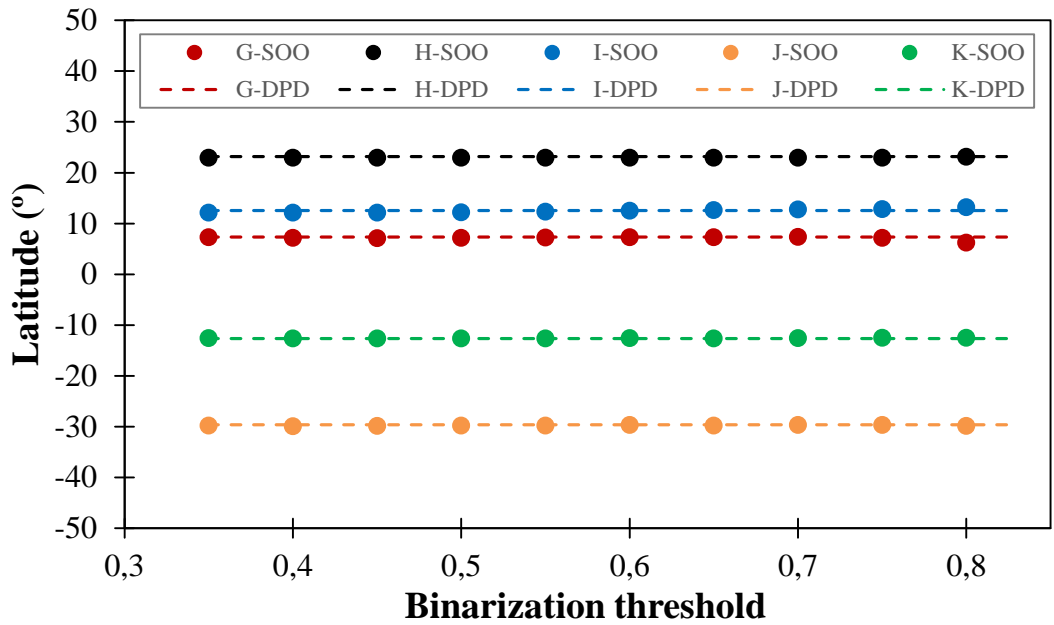

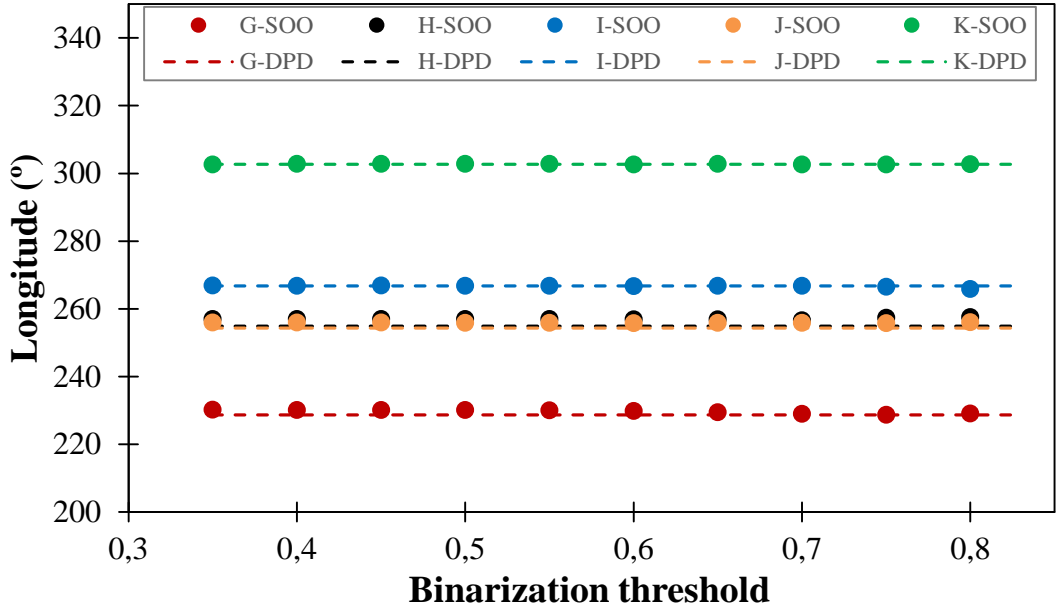

Figure 5. Comparison of the determination of areas (top panel), heliographic latitudes (middle panel) and longitudes (bottom panel) of the sunspot groups recorded by DPD on 15 October 2012 as a function of different binarization thresholds selected in Soonspot. Curves and symbols represents measurements obtained from Soonspot and dashed horizontal lines the data provided by DPD. Red, black, blue, orange, and green colors represent group G, H, I, J, and K recorded by DPD on 15 October 2012, respectively. Areas are given in millionths of solar hemisphere and heliographic latitudes in degrees.

## 5. Conclusions

We present a program for calculating heliographic coordinates and areas of sunspots from digitized images. The software uses three very simple graphical interfaces. In the first one, sunspots are selected. Here, the images can be rotated without quality loss in order to correct its orientation, the solar limb is set with three simple mouse clicks and a binarization threshold must be selected with the aim to measure correctly the sunspot area and coordinates according to the features of the solar drawing or photograph. In the second one, data about the sunspots like the name, area and position in pixel units are provided. In the third interface, after introducing the observation date, information referred to names, areas and heliographic coordinates are shown about sunspots. These data can be saved in a .txt or .xlsx file.

Soonspot: Software to Determine Areas and Sunspot Positions

The program has been checked using modern solar images of the DPD. These results show a very good agreement between the heliographic coordinates determined using Soonspot and data provided by the DPD for the same images. Differences are greater in the case of areas, due to the difficulty in its determination. Furthermore, the program has been used in the analysis of the drawings performed by Hevelius in the 17$^{th}$ century. We show that the number of sunspots observed by Hevelius in the northern hemisphere is slightly greater than the southern. The average value of the umbrae-penumbrae ratio calculated in this work regarding all the sunspots recorded by Hevelius is compatible with values of that parameter obtained in other studies. In addition, we have studied the relative size distribution of the umbral and total group area and they are distributed lognormally such as Bogdan *et al*. (1988) showed using more recent sunspot observations recorded at the Mount Wilson Observatory. Regarding the measurements of heliographic latitudes and longitudes, we can see that the binarization threshold does not have a significant influence in the calculation. Considering the observation made at DPD on 15 October 2012 and averaging all the values obtained in latitudes and longitudes for the same groups with different thresholds, the greatest differences were obtained for groups G for latitudes and H for longitudes with a value equal to 2.5% and 0.8 % with respect to the value provided by DPD, respectively.

**Acknowledgements**

This research was supported by the Economy and Infrastructure Counselling of the Junta of Extremadura through project IB16127 and grants GR18081 (co-financed by the European Regional Development Fund) and by the Ministerio de Economía y Competitividad of the Spanish Government (CGL2017-87917-P). All the historical materials used in this work were consulted at the Biblioteca del Real Observatorio de la Armada (San Fernando, Spain).

**Disclosure of Potential Conflicts of Interest** The authors declare that they have no conflicts of interest.

**References**

Abarbanell, C., Wöhl, H.: 1981, Solar rotation velocity as determined from sunspot drawings of J. Hevelius in the 17$^{th}$ century, *Solar Phys.* **70**, 197-203.

Arlt, R.: 2009, The Butterfly Diagram in the Eighteenth Century, *Solar Phys.* **255**, 143. DOI: 10.1007/s11207-008-9306-5.

Barata, T., Carvalho, S., Dorotovič, I., Pinheiro, J.G., Garcia, A., Fernandes, J., Lourenço, A.M.: 2018, Software tool for automatic detection of solar plages in the Coimbra Observatory spectroheliograms, *Astron. Comput.* **24**, 70. DOI: 10.1016/j.ascom.2018.06.003.

Bogdan, T.J., Gilman, P.A., Lerche, I., Howard, R.: 1988, Distribution of sunspot umbral areas: 1917-1982, ApJ 327, 451. DOI: 10.1086/166206.

Çakmak, H.: 2014, Computer-aided measurement of the heliographic coordinates of sunspot groups, *Exp. Astron.* **38**, 77. DOI: 10.1007/s10686-014-9410-5.

Carrasco, V.M.S., Vaquero, J.M., Gallego, M.C., Sánchez-Bajo, F.: 2016, A Normalized Sunspot-Area Series Starting in 1832: An Update, *Solar Phys.* **291**, 2931. DOI: 10.1007/s11207-016-0943-9.

Carrasco, V.M.S., García-Romero, J.M., Vaquero, J.M., Rodríguez, P.G., Gallego, M.C., Lefèvre, L.: 2018, The Umbra–Penumbra Area Ratio of Sunspots During the Maunder Minimum, *ApJ* **865**, 88. DOI: 10.3847/1538-4357/aad9f6.

Carrasco, V.M.S., Vaquero, J.M., Gallego, M.C., Muñoz-Jaramillo, A., de Toma, G., Galaviz, P., Arlt, R., Senthamizh Pavai, V., Sánchez-Bajo, F., Villalba Álvarez, J., Gómez, J.M., Sunspot characteristics at the onset of the Maunder Minimum based 1 on the observations of Hevelius, *ApJ* 886, 18. DOI: 10.3847/1538-4357/ab4ade.

Cristo, A., Vaquero, J.M., Sánchez-Bajo, F.: 2011, HSUNSPOTS: A tool for the analysis of historical sunspot drawings, *J. Atmos. Sol.-Terr. Phys.* **73**, 187-190. DOI: 10.1016/j.jastp.2009.12.010.

Galaviz, P. 2018, Determinación y análisis de regiones activas solares en los últimos siglos, PhD Thesis (Mérida, University of Extremadura).

Galaviz, P., Vaquero, J.M., Gallego, M.C., Sánchez-Bajo, F.: 2016, A small collection of sunspot drawings made in the Royal Astronomical Observatory of the Spanish Navy in 1884, *Adv. Space Res.* **58**, 2247. DOI: 10.1016/j.asr.2016.08.013.

Soonspot: Software to Determine Areas and Sunspot Positions

Galilei, G., Scheiner, C.: 2010, On Sunspots, University of Chicago Press, Chicago.

Hevelius, J.: 1647, Selenographia: sive lunae description, Hünefeld, Danzig.

Hoyt, D.V., Schatten, K.H.: 1998, Group Sunspot Numbers: A New Solar Activity Reconstruction, *Solar Phys.* **179**, 189. DOI: 10.1023/A:1005007527816.

Hrzina, D., Rosa, D., Hanslmeier, A., Ruzdjak, V., Brajsa, R.: 2007, Sungrabber – Software for measurements on solar synoptic images, *Cent. Eur. Astrophys. Bull.* **31**, 273-279.

Karoff, C., Jørgensen, C.S., Senthamizh Pavai, V., Arlt, R.: 2019, Christian Horrebow's Sunspot Observations – II. Construction of a Record of Sunspot Positions, *Solar Phys.* **294**, 78. DOI: 10.1007/s11207-019-1466-y.

Meadows, P.: 2002, The measurement of sunspot area. *J. Br. Astron. Ass.* **112**, 353.

Meeus, J.: 1991, Astronomical Algorithms, Willmann-Bell, Richmond.

Muñoz-Jaramillo A., Vaquero J.M.: 2018, Visualization of the challenges and limitations of the long-term sunspot number record, *Nat. Astron.* **3**, 205, DOI: 10.1038/s41550-018-0638-2.

Ribes, J.C., Nesme-Ribes, E.: 1993, The solar sunspot cycle in the Maunder minimum AD 1645 to AD 1715, *Astron. Astrophys.* **276**, 549-563.

Sánchez-Bajo, F., Vaquero, J.M., Gallego, M.C.: 2010, Solar Rotation during the Period 1847-1849, *Solar Phys.* **261**, 1. DOI: 10.1007/s11207-009-9469-8.

Vaquero, J.M., Vázquez, M.: 2009, The Sun Recorded Through History, Springer, Berlin.

Vaquero, J.M., Gallego, M.C., García, J.A.: 2002, A 250-year cycle in naked-eye observations of sunspots, *Geophys. Res. Lett.* **29**, 1997. DOI: 10.1029/2002GL014782.

Vaquero, J.M., Gordillo, A., Gallego, M.C., Sánchez-Bajo, F., García, J.A.: 2005, The umbra-penumbra area ratio of sunspots from the de la Rue data, *Obs.* **125**, 152.

Vaquero, J.M., Svalgaard, L., Carrasco, V.M.S., Clette, F., Lefèvre, L., Gallego, M.C., Arlt, R., Aparicio, A.J.P., Richard, J.-G., Howe, R.: 2016, A Revised Collection of Sunspot Group Numbers, *Solar Phys.* **291**, 3061. DOI: 10.1007/s11207-016-0982-2.

Vokhmyanin, M.V., Zolotova, N.V.: 2018, Sunspot Positions and Areas from Observations by Galileo Galilei, *Solar Phys*. **293**, 31. DOI: 10.1007/s11207-018-1245-1.

Willis, D., Stephenson, F.R.: 2001, Solar and auroral evidence for an intense recurrent geomagnetic storm during December in AD 1128, *Ann. Geophys*. **19**, 289. DOI: 10.5194/angeo-19-289-2001.

Yau, K.K.C., Stephenson, F.R.: 1988, A revised catalogue of Far Eastern observations of sunspots (165 BC to AD 1918), *Quart. J. Roy. Astron. Soc*. **29**, 175.